\documentclass[aps,prl,reprint,amsmath,amssymb,superscriptaddress,longbibliography,floatfix]{revtex4-2}

\usepackage{amsmath,amssymb,mathtools}
\usepackage{bm}
\usepackage{graphicx}
\usepackage{etoolbox}
\usepackage{xcolor}
\usepackage[colorlinks=true,citecolor=blue,linkcolor=blue,urlcolor=blue]{hyperref}

\begin{document}

\title{Momentum-dependent phonon effects on spin-fluctuation $d$-wave pairing in bulk La$_6$Ni$_5$O$_{12}$}

\author{Dexi Shao}
\thanks{Dexi Shao and Zhiming Pan contributed equally to this work.}
\affiliation{School of Physics and Hangzhou Key Laboratory of Quantum Matter, Hangzhou Normal University, Hangzhou 311121, China}
\author{Zhiming Pan}
\email{panzhiming@xmu.edu.cn}
\affiliation{Department of Physics, Xiamen University, Xiamen 361005, Fujian, China}
\author{Ming Zhang}
\affiliation{Zhejiang Key Laboratory of Quantum State Control and Optical Field Manipulation, Department of Physics, Zhejiang Sci-Tech University, Hangzhou 310018, China}
\author{Yupeng Li}
\affiliation{School of Physics and Hangzhou Key Laboratory of Quantum Matter, Hangzhou Normal University, Hangzhou 311121, China}
\author{Chen Lu}
\email{luchen@hznu.edu.cn}
\affiliation{School of Physics and Hangzhou Key Laboratory of Quantum Matter, Hangzhou Normal University, Hangzhou 311121, China}

\begin{abstract}
We study the effect of electron-phonon coupling on spin-fluctuation-mediated pairing in bulk La$_6$Ni$_5$O$_{12}$.
A five-layer Hubbard model treated within the random-phase approximation yields a leading $d_{x^2-y^2}$ pairing state.
Two phenomenological couplings motivated by oxygen breathing modes favor $s$-wave pairing when considered alone, while their attractive scattering between opposite-sign regions of the spin-fluctuation-induced gap suppresses $d$-wave pairing.
Electronic charge screening strongly reduces this suppression.
Comparing the $s$- and $d$-wave pairing strengths with and without screening shows that screening shifts their boundary in favor of $d$-wave pairing.
Momentum-resolved projections locate the reduction of the unfavorable scattering processes.
At the representative electronic interaction, $d$-wave pairing remains dominant.
\end{abstract}

\maketitle

\textit{Introduction.---}
The early search for superconductivity (SC) in nickelates was motivated by their structural and electronic similarities to cuprates~\cite{Anisimov1999,LeePickett2004}. 
A relevant candidate is the reduced Ruddlesden-Popper series $R_{n+1}$Ni$_n$O$_{2n+2}$, where $R$ denotes a rare-earth element and $n$ represents the basis layer number within a block.
SC was first discovered in hole-doped infinite-layer thin films~\cite{Li2019}.
Subsequent studies established superconducting domes in Nd- and Pr-based compounds~\cite{Li2020Dome,Zeng2020Dome,Osada2020PrPhase} and SC in La-based films~\cite{Osada2021LaSr,Zeng2022LaCa}. 
The realization of SC in quintuple-layer Nd$_6$Ni$_5$O$_{12}$ films extended this family to $n=5$~\cite{Pan2022}.
More recently, bulk SC was reported in La$_6$Ni$_5$O$_{12}$ at ambient pressure,
with an onset $T_c$ near $18$~K~\cite{Shi2026}.

The low-energy descriptions and SC of these nickelates are dominated by the Ni $d_{x^2-y^2}$ orbital within each elementary NiO$_2$ layer~\cite{BotanaNorman2020,Nomura2019,Karp2020,Hepting2020,LaBollita2022many}.
In infinite-layer nickelates, the observed dispersive magnetic excitations~\cite{Lu2021Magnetic} and signatures of nodal pairing~\cite{Cheng2024dwave,Harvey2025Nodal} suggest unconventional pairing.
For infinite-layer compounds, theoretical studies identify antiferromagnetic (AFM) spin fluctuations as a possible pairing mechanism and favor $d$-wave  pairing~\cite{Wu2020Robust,Sakakibara2020,Kitatani2020,
Worm2024Spin}.
Multiorbital studies have also examined the role of Ni $d_{z^2}$ states~\cite{Kreisel2022,Lechermann2022Flat}.
Correlated-model calculations find similar $d$-wave
pairing tendencies in finite-layer
nickelates~\cite{Worm2022,Hausoel2025}.
For La$_6$Ni$_5$O$_{12}$, recent variational Monte Carlo
calculations also favor a $d$-wave state~\cite{Wang2026}.

Electron-phonon coupling can influence SC when electronic correlations provide the dominant pairing interaction.
Early random-phase approximation (RPA) studies showed that different phonon interactions can enhance or suppress spin-fluctuation-mediated $d$-wave pairing depending on their momentum dependence~\cite{Bulut1996}.
Cooperative effects have also been found in other settings.
Forward-focused interfacial phonons can enhance spin-fluctuation-mediated pairing in FeSe/SrTiO$_3$~\cite{Rademaker2021}, 
while out-of-plane phonons have been proposed to enhance $s_\pm$ pairing in bilayer La$_3$Ni$_2$O$_7$~\cite{Zhan2025Cooperation}.

Assessing the phonon contribution requires understanding the momentum and orbital structure of the electron-phonon interaction.
In cuprates, phonon symmetry and charge screening give rise to strongly anisotropic couplings to oxygen modes~\cite{Devereaux2004,Johnston2010}.
On the kagome lattice, the sublattice composition of Fermi-surface states provides another source of momentum dependence in the effective phonon-mediated interaction~\cite{Wu2024EPC}.
For the five-layer nickelate La$_6$Ni$_5$O$_{12}$, first-principles calculations find appreciable coupling to oxygen breathing distortions associated with charge and bond disproportionation~\cite{CarrascoAlvarez2024}.

In this Letter, we study the effect of oxygen breathing modes on spin-fluctuation-mediated pairing in La$_6$Ni$_5$O$_{12}$.
Phenomenological electron-phonon couplings for these modes are incorporated into a five-layer Hubbard model.
We evaluate the pairing kernel within the RPA using either bare electron-phonon vertices or vertices renormalized by the RPA charge response.
This unified treatment reveals the momentum-space structure of the phonon contribution and identifies the scattering processes detrimental to spin-fluctuation-induced $d_{x^2-y^2}$ pairing.
We find that charge screening preferentially weakens these processes.
As a result, $d$-wave pairing remains dominant over a wider range of electron-phonon coupling strengths.

\textit{Five-layer electronic structure.---}
La$_6$Ni$_5$O$_{12}$ consists of five-layer NiO$_2$ blocks separated by fluorite-type La$_2$O$_2$ slabs~\cite{LaBollita2022,Shi2026}.
Nominal charge counting gives a Ni valence of $+1.2$ with a $3d^{8.8}$ configuration, corresponding to $0.2$ holes per Ni.
To describe the low-energy Ni $3d_{x^2-y^2}$ electrons, we use the following five-layer Hubbard Hamiltonian:
\begin{equation}
H_{\mathrm e}=\sum_{\mathbf{k}\sigma}
\mathbf{c}_{\mathbf{k}\sigma}^{\dagger}
h_{\mathrm{TB}}(\mathbf{k})
\mathbf{c}_{\mathbf{k}\sigma}
+U\sum_{ia}n_{ia\uparrow}n_{ia\downarrow}.
\label{eq:He}
\end{equation}
Here, the $a$th component of the five-component layer spinor $\mathbf{c}_{\mathbf{k}\sigma}$ is $c_{\mathbf{k}a\sigma}$, which annihilates an electron with momentum $\mathbf{k}$ and spin $\sigma$ on layer $a$. The outer layers ($a=1,5$) are symmetry equivalent, as are the inner layers ($a=2,4$). The tight-binding matrix follows Ref.~\cite{Wang2026} and has diagonal elements
\begin{equation}
\begin{aligned}
&[h_{\mathrm{TB}}(\mathbf{k})]_{aa}
=\epsilon_a+2t_{1a}(\cos k_x+\cos k_y) \\
+& 4t_{2a}\cos k_x\cos k_y +2t_{3a}(\cos 2k_x+\cos 2k_y).
\end{aligned}
\label{eq:tbdiag}
\end{equation}
Adjacent layers are coupled by $[h_{\mathrm{TB}}]_{a,a+1}=[h_{\mathrm{TB}}]_{a+1,a}=-2t_{\perp,a}(\cos k_x-\cos k_y)^2$.
The details of the parameters are given in Sec.~S1 of Supplemental Material (SM)~\cite{SupplementalMaterial}.

\begin{figure}[!t]
\centering
\includegraphics[width=0.9\columnwidth]{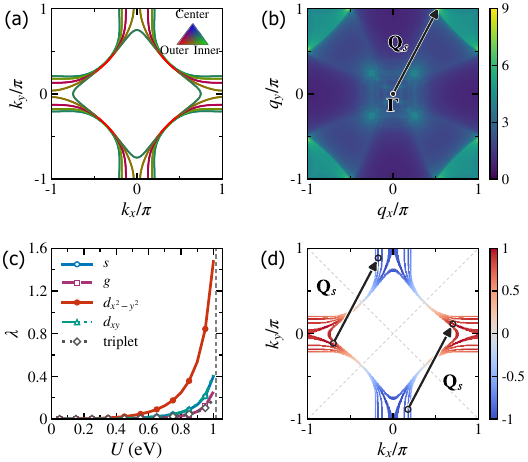}
\caption{Spin-fluctuation-mediated pairing. 
(a) Fermi surface colored by outer-, inner-, and center-layer character. 
(b) Largest eigenvalue of the spin susceptibility at $U=0.9$~eV, with critical $U_c\simeq1.016$~eV. 
The arrow from $\Gamma$ points to the magnetic maximum at $\mathbf{Q}_s$. 
(c) Pairing strengths versus $U$; 
the dashed line denotes $U_c$. 
(d) RPA $d_{x^2-y^2}$ gap at $U=0.9$~eV, normalized to $\max|\Delta|=1$. 
Arrows indicate representative $\mathbf{Q}_s$ transfers.}
\label{Fig1}
\end{figure}

Fig.~\ref{Fig1}(a) shows that the five bands give rise to closely spaced Fermi-surface sheets with mixed layer character, consisting of a central electronlike pocket and four surrounding holelike pockets. We restrict pairing to these five Ni-derived bands.

\textit{Spin-fluctuation-mediated pairing.---}
In the RPA framework, we analyze spin fluctuations and the associated superconducting instability~\cite{Kuroki2008RPA,Graser2009RPA,Lu2018Triangular,Lu2022SO4,Zhang2024Trilayer};
the details are given in Sec.~S1 of SM.
The bare susceptibility includes intraband and interband
particle-hole excitations.
Fermi-surface nesting can enhance this response at finite wave vectors.
Calculations found the largest eigenvalue of the spin susceptibility $\chi^s$ peaks at $\mathbf{Q}_s\simeq (0.53\pi,\pi)$ and its symmetry-related momenta, as shown in Fig.~\ref{Fig1}(b).
Spin susceptibility diverges at a critical interaction about $U_c\simeq1.016$~eV, 
signaling an instability toward magnetic order at $\mathbf{Q}_s$.
We study pairing in the paramagnetic regime with $U<U_c$.

The effective pairing interaction contains both contributions from charge and spin fluctuations.
After projection onto the band basis, we solve the symmetrized linearized gap equation for the pairing kernel $K_{\mathrm{el}}$ in each symmetry channel $\alpha$ and obtain the pairing eigenvalues $\lambda_\alpha$.
Figure~\ref{Fig1}(c) shows $\lambda_\alpha$ in the relevant symmetry channels as a function of $U$.
In the weak-coupling regime, superconducting transition temperature scales roughly as $T_{c,\alpha}\propto \omega_c e^{-1/\lambda_\alpha}$ for a common energy cutoff $\omega_c$.
The channel with the largest $\lambda_\alpha$ gives the leading pairing instability.

The pairing strengths of the four even-parity channels and the triplet sector as a function of $U$ are summarized in Fig.~\ref{Fig1}(c). 
For $U<U_c$, the $d_{x^2-y^2}$ channel has the largest pairing strength and gives the leading pairing instability. 
Spin fluctuations near $\mathbf{Q}_s$ mediate strong repulsive pair scattering in the singlet channel. 
This scattering favors opposite gap signs on the Fermi-surface regions connected by $\mathbf{Q}_s$, 
consistent with the $d_{x^2-y^2}$ gap shown in Fig.~\ref{Fig1}(d).

We further examine the layer dependence of magnetic order and pairing.
The leading magnetic mode is concentrated in the middle three layers, whereas the outer layers carry the largest projected gap weights. Details of the analysis are provided in Sec.~S1 of SM.

\textit{Electron-phonon coupling.---}
Frozen-phonon calculations for metallic La$_6$Ni$_5$O$_{12}$ find that two in-plane oxygen breathing distortions associated with bond disproportionation split the bands near the Fermi level~\cite{CarrascoAlvarez2024}. 
We take their wave vectors and energies as inputs to two model couplings. 
For mode A, the momentum envelope $F_A(\mathbf{q})$ peaks at $\mathbf{Q}_A=(\pi,\pi)$ with phonon energy $\Omega_A=69$~meV. 
For mode B, $F_B(\mathbf{q})$ peaks at $\mathbf{Q}_B=(2\pi/3,2\pi/3)$ and the three symmetry-related points, with $\Omega_B=66$~meV. 
The periodic envelopes have unit peak height and a common model width $\sigma=\pi/6$ (see Sec.~S2 in SM).

This interaction supplies additional attractive pair scattering with the momentum dependence specified by $F_\mu(\mathbf{q})$.
We consider each mode separately and denote the sum of their independent exchange interactions by AB. 
Each included mode has the coupling amplitude $g_\mu=g$. 
As a local reference (ISO), we retain both phonon energies and set $F_A=F_B=1$. 
The bare ISO interaction then couples on-site pairs within each layer. 
For a $B_{1g}$ gap, the on-site pair amplitude cancels on every layer because the gap changes sign under $k_x\leftrightarrow k_y$ while the layer probabilities are even. 
The bare ISO kernel vanishes throughout the $B_{1g}$ sector (SM, Sec.~S2).

\begin{figure}[!t]
\centering
\includegraphics[width=0.9\columnwidth]{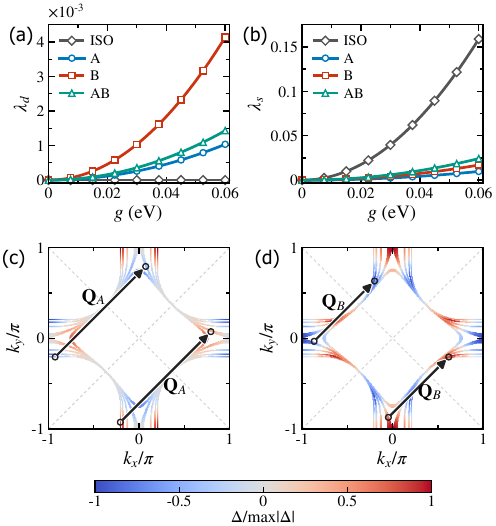}
\caption{Phonon-mediated pairing at $U=0$. 
(a,b) Leading pairing strengths in the $d_{x^2-y^2}$ and $s$ channels, respectively.
The local (ISO) interaction has zero $d$-wave pairing strength. 
(c,d) $d$-wave gaps for A and B at $g=0.06$~eV, normalized to $\max|\Delta|=1$. 
Arrows mark representative $\mathbf{Q}_A$ and $\mathbf{Q}_B$ transfers.}
\label{Fig2}
\end{figure}

We first study pairing driven by electron-phonon coupling at $U=0$.
These electron-phonon vertices yield the bare phonon-mediated interaction $W_{\mathrm{ph},\mu}^{(\mathrm{bare})}(\mathbf{q})=-2g_\mu^2F_\mu(\mathbf{q})I/\Omega_\mu$, where $\mu=\mathrm{A},\mathrm{B}$ and $I$ is the identity matrix. Modes A and B mediate attractive pair scattering near $\mathbf{Q}_A$ and $\mathbf{Q}_B$, respectively, connecting Fermi-surface regions near the $k_x$ and $k_y$ axes.
Such scattering favors equal gap signs on the connected regions~\cite{Bulut1996,Graser2009RPA}.
As summarized in Fig.~\ref{Fig2}(a,b), 
all four models favor $s$-wave pairing, while the momentum-dependent interactions in A, B, and AB also support subleading $d_{x^2-y^2}$ pairing.
Mode B gives the largest $d$-wave pairing strength, which is reduced when mode A is added.
The local ISO interaction instead gives the strongest $s$-wave pairing and vanishes in the $d$-wave channel.

The gap structures in Fig.~\ref{Fig2}(c,d) show how the attractive scattering is accommodated within the $B_{1g}$ sector.
For both modes, the gap changes sign between neighboring Fermi-surface sheets, allowing the regions connected by the indicated $\mathbf{Q}_A$ or $\mathbf{Q}_B$ transfers to carry the same sign.
The different momentum transfers favor distinct distributions of gap signs and amplitudes for A and B.
Both phonon-induced gaps differ from the spin-fluctuation-induced gap in Fig.~\ref{Fig1}(d), where repulsive scattering near $\mathbf{Q}_s$ favors opposite signs on the connected regions.
The momentum structure of the pairing interaction determines different gap structures even within the same $B_{1g}$ symmetry.

\textit{Charge screening and pairing competition.---}
We now consider how electron-phonon coupling modifies spin-fluctuation-mediated pairing at finite $U$.
We first retain the electron-phonon vertices $W_{\mathrm{ph},\mu}^{(\mathrm{bare})}(\mathbf{q})$ used in the $U=0$ limit and combine it with the spin-fluctuation-mediated pairing interaction at the chosen $U$. 
Solving the full eigenproblem determines the resulting pairing strengths $\lambda_{\nu,\alpha}^{(\mathrm{bare})}$.
We compare these results with the values $\lambda_\alpha$ in Fig.~\ref{Fig1}(c), obtained at the same $U=0.9$eV without phonon exchange.

The electronic interaction will also renormalize the phonon-mediated vertices through the charge response to lattice displacements.
This screening can be included within static RPA~\cite{Wu2024EPC}
\begin{equation}
\begin{aligned}
W_{\mathrm{ph},\mu}^{(\mathrm{scr})}(\mathbf{q};U)
&=-\frac{2g_\mu^2}{\Omega_\mu}F_\mu(\mathbf{q})
S(\mathbf{q};U)S^T(\mathbf{q};U),\\
S(\mathbf{q};U)
&=\bigl[I+UC_0(\mathbf{q})\bigr]^{-1},
\end{aligned}
\label{eq:Wph}
\end{equation}
where $[C_0(\mathbf{q})]_{ab}=\chi^0_{aa,bb}(\mathbf{q})$ is the noninteracting layer-density susceptibility obtained from $h_{\mathrm{TB}}$.
At each $U$ and $g$, the electronic pairing interaction and phonon propagator are held fixed when incorporating this vertex correction.
The phonon-induced changes are measured by $\Delta\lambda_{\nu,\alpha}^{(\mathrm{x})}=\lambda_{\nu,\alpha}^{(\mathrm{x})}-\lambda_\alpha$, where $\nu$ labels the phonon model, $\alpha=s,d$, and $\mathrm{x}=\mathrm{bare},\mathrm{scr}$ denotes the original and renormalized vertices, respectively.

\begin{figure}[!t]
\centering
\includegraphics[width=0.9\columnwidth]{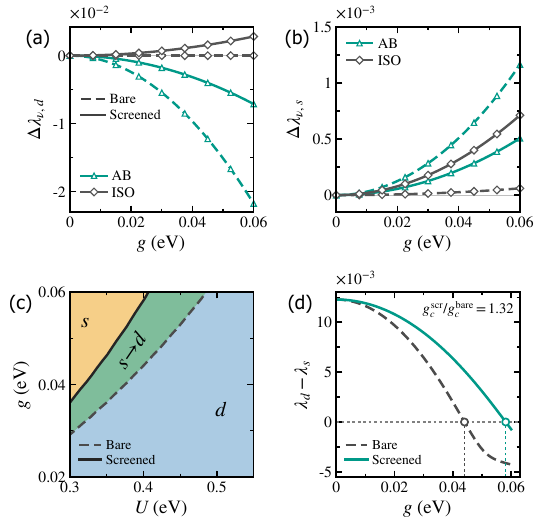}
\caption{Interplay between phonon effect and electron correlation.
(a,b) Phonon-induced changes in the $d$- and $s$-wave pairing strengths for AB and ISO at $U=0.9$~eV.
(c) AB boundaries $g_c(U)$ defined by $\lambda_s=\lambda_d$.
The yellow and blue regions favor $s$- and $d$-wave pairing, respectively, in both treatments.
In the green region, the original vertices favor $s$-wave pairing and the screened vertices favor $d$-wave pairing.
(d) Eigenvalue difference $\lambda_d-\lambda_s$ at $U=0.4$~eV, with circles marking $g_c$.
Dashed and solid curves correspond to original and electronically screened vertices, respectively.}
\label{Fig3}
\end{figure}

The results are summarized in Fig.~\ref{Fig3}(a,b),
which show the phonon-induced changes in the pairing strengths at $U=0.9$~eV.
For AB, increasing $g$ suppresses $d$-wave pairing and enhances $s$-wave pairing.
Electronic screening weakens both effects and suppresses the $d$-wave pairing strength, as depicted in Fig.~\ref{Fig3}(a).
The local ISO interaction leaves the $d$-wave strength unchanged with the original vertices, while screening produces a small enhancement.
Screening also increases the ISO contribution to $s$-wave pairing, in contrast to its effect on AB; see Fig.~\ref{Fig3}(b).
At fixed $U$, the separation between the electronic $s$- and $d$-wave pairing strengths is sufficiently large that $d$-wave pairing remains dominant throughout $0\leq g\leq0.06$~eV;
see Sec.~S3 in SM for details.

The interplay between electron-phonon and electronic correlation determines the dominant pairing channel, as depicted in Fig.~\ref{Fig3}(c).
Larger Hubbard $U$ favors $d$-wave pairing, while stronger electron-phonon coupling $g$ prefers $s$-wave pairing.
The boundaries between them rise with $U$.
Taken into the effect of screening will shift the quantitative boundary, while keeping the main feature invariant.

The shift of the boundary is explicitly illustrated  at $U=0.4$~eV in Fig.~\ref{Fig3}(d).
The difference $\lambda_d-\lambda_s$ is positive when $d$-wave pairing dominates and negative when $s$-wave pairing dominates.
Increasing $g$ reduces this difference, while the reduction is weaker when taken into electronic screening.
The zero crossing $g_c$ moves to larger value.
The interval between these crossings corresponds to the green region in Fig.~\ref{Fig3}(c).

\textit{Momentum-resolved phonon contributions.---}
To identify the scattering processes that change the $d$-wave pairing strength, we project the phonon kernel onto the normalized $d$-wave eigenstate $\psi_d$ of the purely electronic kernel $K_{\mathrm{el}}$,
\begin{equation}
C_{\nu,d}^{(\mathrm{x})}(g)
=-\bigl\langle\psi_d\big|K_{\mathrm{ph},\nu}^{(\mathrm{x})}(g)\big|\psi_d\bigr\rangle.
\label{eq:C}
\end{equation}
For a nondegenerate $\psi_d$, the eigenvalue shift is $\Delta\lambda_{\nu,d}^{(\mathrm{x})}=C_{\nu,d}^{(\mathrm{x})}+O(g^4)$ since $K_{\mathrm{ph},\nu}^{(\mathrm{x})}\propto g^2$. 
The screened projections reproduce the shifts from diagonalization at $U=0.9$~eV; see Sec.~S3 in SM. 
Each breathing-mode kernel has a positive pairing projection on a phonon-induced $d$-wave gap and a negative projection on $\psi_d$. 
The sign depends on the gap structure within $B_{1g}$.

\begin{figure}[!t]
\centering
\includegraphics[width=0.9\columnwidth]{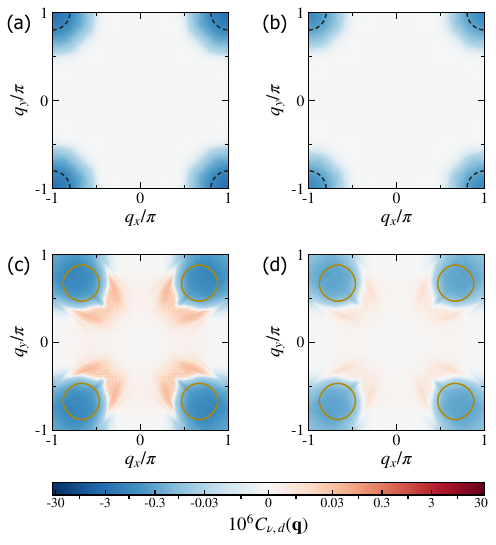}
\caption{Momentum-resolved phonon projections on the spin-fluctuation-induced $d$-wave state at $U=0.9$~eV and $g=0.06$~eV. (a,b) Mode A with bare and screened vertices. (c,d) Mode B with bare and screened vertices. The four panels share a symmetric logarithmic color scale. Dashed and solid contours mark $F_A=0.5$ and $F_B=0.5$, respectively.}
\label{Fig4}
\end{figure}

We group the contributions to Eq.~\eqref{eq:C} by transferred momentum. 
For singlet scattering from $(\mathbf k,-\mathbf k)$ to $(\mathbf k',-\mathbf k')$, we include both $\mathbf{q}=\mathbf k-\mathbf k'$ and $\mathbf{q}=\mathbf k+\mathbf k'$, with the band matrix elements and energy-shell weights. 
The resulting map satisfies
\begin{equation}
\sum_{\mathbf{q}}C_{\nu,d}^{(\mathrm{x})}(\mathbf{q})=C_{\nu,d}^{(\mathrm{x})}.
\label{eq:qsum}
\end{equation}
An attractive transfer contributes positively when the two gap values have the same sign and negatively when they have opposite signs.
The results are summarized in Fig.~\ref{Fig4}.

\begin{table}[b]
\caption{Phonon projections on $\psi_d$ at $U=0.9$~eV and $g=0.06$~eV. For a negative bare projection, $R=100(1-|C_d^{\mathrm{scr}}/C_d^{\mathrm{bare}}|)$ is its percentage reduction in magnitude. The bare ISO projection is zero.}

\label{tab:phonon_screening}
\begin{ruledtabular}
\begin{tabular}{lrrr}
Model & $10^3C_d^{\mathrm{bare}}$ & $10^3C_d^{\mathrm{scr}}$ & $R$ (\%)\\
ISO & $0$ & $+2.82$ & ---\\
A & $-7.97$ & $-2.55$ & $68.0$\\
B & $-13.81$ & $-4.57$ & $66.9$\\
AB & $-21.78$ & $-7.12$ & $67.3$
\end{tabular}
\end{ruledtabular}
\end{table}

For mode A, the largest negative contribution comes from $\mathbf{q}=(\pi,\pi)$; see Fig.~\ref{Fig4}(a). 
For B, it lies on the zone diagonals near $(|q_x|,|q_y|)=(0.77\pi,0.77\pi)$, displaced from the envelope peaks at $(2\pi/3,2\pi/3)$ and equivalent points; see Fig.~\ref{Fig4}(c). 
The phonon envelope sets the available scattering weight, while the gap signs and layer content of the connected states determine its contribution to pairing. 
As shown in Fig.~\ref{Fig4}(b,d), screening attenuates the dominant negative contributions for both modes. 
Their momentum sums remain negative, with magnitudes reduced by about two thirds; see Tab.~\ref{tab:phonon_screening}.

For the local reference, same-sign and opposite-sign scattering cancel in the bare momentum sum. 
Screening reduces both contributions, but opposite-sign pairs place more weight in layer-density modes with a larger charge response and are attenuated more strongly. 
The imbalance gives a small positive projection on $\psi_d$ (Table~\ref{tab:phonon_screening}). 
Replacing the screening factor by its momentum average restores the cancellation for ISO while still reducing the suppression for A and B; see Sec.~S4 in SM.

\textit{Discussion.---}
In summary, we found both oxygen breathing modes could suppress spin-fluctuation-mediated $d$-wave pairing.
This suppression comes from attractive scattering between Fermi-surface regions where the gap has opposite signs.
Electronic screening of the electron-phonon vertices weakens this scattering.
With both modes present, screening shifts the crossover from $d$-wave to $s$-wave pairing to larger electron-phonon coupling $g$.

In our calculation, the spin-fluctuation-induced $d$-wave gap carries more weight on each outer NiO$_2$ layer than on the central layer.
In the five-layer cuprate Hg-1245, the inner and outer planes differ in both Cu coordination and hole density~\cite{Kotegawa2004,Kotegawa2001}.
In La$_6$Ni$_5$O$_{12}$, Ni has square-planar coordination in all five layers, and the calculated orbital occupations vary only weakly from layer to layer~\cite{LaBollita2022}.

We use the same local electron-phonon coupling on all five layers before screening.
Holding the electronic pairing interaction fixed isolates the direct effect of phonon exchange on pairing.
Functional renormalization group calculations for trilayer nickelates show that layer-dependent electron-phonon couplings can also modify magnetic fluctuations~\cite{Baum2026ModeSelective}.
A quantitative estimate of the phonon-induced shift in $T_c$ for La$_6$Ni$_5$O$_{12}$ would require the momentum and frequency dependence of its electron-phonon couplings.

\textit{Acknowledgments.---}
This work was supported by the National Natural Science Foundation of China (Grant Nos.~12304180, 12504219, 12404154, and 12204138), the Zhejiang Provincial Natural Science Foundation of China (Grant Nos.~ZCLQN25A0402 and LMS26A040010), startup funding from Xiamen University, the Beijing National Laboratory for Condensed Matter Physics (Grant No.~2025BNLCMPKF013), the HZNU Scientific Research and Innovation Team Project (Grant No.~TD2025013), the Hangzhou Leading Youth Innovation and Entrepreneurship Team Project (Grant No.~TD2024005), and the Dengfeng Research Grant Program from Hangzhou Normal University (Grant No.~TD2026008).

\textit{Data availability.---}
The numerical data and calculation codes supporting this work are available from the corresponding authors upon reasonable request.

\bibliography{references}

\end{document}